# ENTERPRISE VALUE, ECONOMIC AND POLICY UNCERTAINTIES: THE CASE OF US AIR CARRIERS

**Bahram Adrangi, PhD**
University of Portland
5000 N. Willamette Blvd.
Portland, Oregon 97203
adrangi@up.edu

**Arjun Chatrath, PhD**
Schulte Professor of Finance
University of Portland
5000 N. Willamette Blvd.
Portland, Oregon 97203
chatrath@up.edu

**Madhuparna Kolay, PhD**
University of Portland
5000 N. Willamette Blvd.
Portland, Oregon 97203
kolay@up.edu

**Kambiz Raffiee, PhD**
University of Nevada, Reno
Reno, Nevada 89557
raffiee@unr.edu

**Key words: enterprise value, economic policy uncertainty, global economic policy uncertainty, current ratio, debt-to-asset ratio, VAR_MIDAS**

# ENTERPRISE VALUE, ECONOMIC AND POLICY UNCERTAINTIES: THE CASE OF US AIR CARRIERS

## ABSTRACT

The enterprise value (EV) is a crucial metric in company valuation as it encompasses not only equity but also assets and liabilities, offering a comprehensive measure of total value, especially for companies with diverse capital structures. The relationship between economic uncertainty and firm value is rooted in economic theory, with early studies dating back to Sandmo's work in 1971 and further elaborated upon by John Kenneth Galbraith in 1977. Subsequent significant events have underscored the pivotal role of uncertainty in the financial and economic realm. Using a VAR-MIDAS methodology, analysis of accumulated impulse responses reveals that the EV of air carrier firms responds heterogeneously to financial and economic uncertainties, suggesting unique coping strategies. Most firms exhibit negative reactions to recessionary risks and economic policy uncertainties. Financial shocks also elicit varied responses, with positive impacts observed on EV in response to increases in the current ratio and operating income after depreciation. However, high debt levels are unfavorably received by the market, leading to negative EV responses to debt-to-asset ratio shocks. Other financial shocks show mixed or indeterminate impacts on EV.

## INTRODUCTION

Enterprise value is an important metric for valuing a company because it takes into account all of the company's assets and liabilities, not just its equity. This makes it a more accurate measure of the company's overall worth, especially when comparing companies with different capital structures.

EV is calculated by adding the market capitalization of the company's equity to the value of its debt, less the value of its cash and cash equivalents. The formula is as follows:

EV = Market capitalization + Debt - Cash and cash equivalents. Market capitalization is the total value of the company's shares, which is calculated by multiplying the number of shares outstanding by the current share price. Debt is the total amount of money that the company owes to its creditors. Cash and cash equivalents are the company's liquid assets, such as cash in the bank and short-term investments.

EV is used by investors, analysts, and bankers to compare the value of different companies. It is also employed to estimate the total value of a company. Analysts calculate the EV/Sales of a company to gauge its attractiveness as an investment. For example, a company with a lower EV/Sales ratio (less than 3) is considered undervalued and a valuable investment target.

In addition to being a more accurate measure of a company's value, EV can also be used to assess a company's financial health. For example, a company with a high debt-to-equity ratio will have a lower EV than a company with a low debt-to-equity ratio. This is because the company with the high debt-to-equity ratio has more liabilities, which reduces its overall value.

In summary EV is an important metric for valuing a company. It is more accurate than market capitalization and can be used to assess a company's financial health. Table 1 presents some of the main reasons that EV is a key financial metric as well as its shortcomings.

Table 1: Pros and Cons of Enterprise Value (EV)

| |
|---|
| Pros |
| It is a more comprehensive measure of a company's value than market capitalization. |
| It can be used to compare companies with different capital structures. |
| It can be used to assess a company's financial health. |
| It can be used to determine the cost of acquiring a company. |
| It can be used to analyze the performance of a company over time. |
| Cons |
| It does not account for intangible assets, such as intellectual property |
| It can be difficult to calculate EV for companies that have complex financial structures. |

Overall, enterprise value is a valuable tool for valuing companies. However, it is important to use it in conjunction with other metrics to get a complete picture of a company's value. The main goal of this research is to investigate the impact of economic and policy uncertainties on EV of a select group of air carrier firms in the US.

Economic uncertainties refer to the unpredictability of economic conditions, which can arise due to various factors, such as changes in government policies, geopolitical tensions, global economic shocks, and natural disasters. Such uncertainties can have a significant impact on the enterprise value of firms, which is the market value of a company's total equity and debt.

The relationship between economic uncertainty and firm value is rooted in economic theory. Early investigations into the influence of economic uncertainty on corporate behavior trace their origins back to the groundbreaking work of Sandmo (1971). Following Galbraith's work, 'The Age of Uncertainty,' in 1977, numerous significant events, covered extensively in both the media and academic circles, underscored the pivotal role of uncertainty in the economic and financial realm. Building upon this foundation, subsequent scholars, including Flacco and Kroetch (1986), Fooladi and Kayhani (1990, 1991), and Adrangi and Raffiee (1999), have delved into diverse facets of how firms respond to uncertainty. These researchers have undertaken comprehensive theoretical inquiries to unravel the intricate ways in which firms may adjust their production processes, pricing strategies, and profit-seeking activities in the face of market risks and uncertainties.

Further pioneering work by Bernanke (1983), Pindyck (1991), and Dixit and Pindyck (1994) lead to the theory of investment under uncertainty. This theory suggests that heightened uncertainty leads to delayed investment decisions, which may affect a firm's EV.

While there is unanimous agreement on the importance of uncertainty, a single, universally accepted definition of the term remains elusive. Furthermore, it wasn't until relatively recently that scholars began to empirically investigate the role of uncertainty in the economy and financial markets. Recent developments in the measurement of economic and economic policy uncertainties were spearheaded by the work of Baker et al. (2013, 2014, 2016) which has played a pivotal role in quantifying and analyzing economic and economic policy uncertainties. Their research has provided indices that provide valuable insights into the dynamics of economic uncertainty, helping policymakers, economists, and investors better understand the intricacies of these critical factors.

As asserted by Baker et al. (2016), in the aftermath of the 2008 global financial crisis, uncertainty surrounding government policies reached its zenith. This was primarily due to the uncertainty businesses and households faced concerning the government's future stance on regulatory frameworks, spending, taxation, monetary policies, and healthcare. The authors contend that this policy uncertainty significantly hindered the recovery from the recession, as businesses and households delayed their decisions regarding investment and consumption expenditures.

Attention to sentiment and uncertainty indicators have led researchers to quantity various aspects of uncertainty. Manela and Moreira (2017) introduced a monthly news-driven metric called NVIX, while Baker et al. (2012a, 2012b, 2013,2014,2016, 2019, 2021) developed several indices designed to capture economic and financial uncertainties. They also devised Twitter-based measurements of economic uncertainty (TEU) and other tools for quantifying market uncertainties.

Economic uncertainties can lead to increased volatility in markets and hinder consumption, investment, production, and employment, ultimately slowing economic growth (Baker et al., 2016; Bloom, Bond & van Reenen, 2007; Bloom, 2009). In recent years, there has been a growing focus in research on quantifying media data to make it more accessible for economic analyses. This trend reflects a recognition of the significant impact media coverage has on shaping societal perceptions and influencing economic development. Known as the "agenda setting effect," this process involves the selection of information and sources, as well as the tone of reporting, which can profoundly shape public understanding of economic uncertainties and risks (Baker et al., 2016, p. 1595). These influencing factors play a critical role in economic dynamics.

The Economic Policy Uncertainty Index (EPU), based on data from ten major US daily newspapers, serves as a tool to measure the impact of media coverage on economic sentiment. Researchers collect articles from digital archives dating back to January 1985, using specific search criteria to identify relevant content related to uncertainty, the economy, and policy. An article is considered relevant if it includes at least one word from three different sets of terms each. These are “uncertainty” or “uncertain”; “economic” or “economy”; and one of the following policy terms: “Congress”, “deficit”, “Federal Reserve”, “legislation”, “regulation”, or “White House” (Baker et al.2016). The frequency of relevant articles determines the trajectory of the index, which is compiled on a monthly basis (Baker, Bloom & Davis, 2012).

The Global Economic Policy Uncertainty (GEPU) Index is an economic indicator developed by economists Bloom et al. (1997) that measures perceptions of uncertainty and predicts economic futures. The GEPU Index is a GDP-weighted average of national Economic Policy Uncertainty (EPU) indices for countries that account for two-thirds of global output. Each national EPU index is based on the frequency of newspaper articles that mention the economy, uncertainty, and policy-related topics. The GEPU Index is constructed by:

1. Renormalizing each national EPU index to a mean of 100 from 1997 to 2015
2. Imputing missing values for some countries using regression
3. Calculating the GEPU Index value for each month as the GDP-weighted average of the national EPU index values

The GEPU Index is consistent with a principal componentanalysis- based (PCA-based) global economic policy uncertainty index, which is positively correlated with global financial market volatility and correlation. This suggests that higher global economic policy uncertainty leads to more volatile and correlated stock markets.

Moreover, Baker, Bloom, and Davis (2016, p. 1604) have expanded their research to develop similar indices for other countries, including Germany, France, Japan, as well as emerging economies like China, India, and Russia. This broader scope allows for a more comprehensive understanding of the global impact of media-driven economic perceptions.

These indicators have gained substantial importance, particularly in the analysis of volatility across various asset categories. Prominent studies in this domain include those by Su et al. (2017, 2018, 2019), Altig et al. (2020), Krol (2014), Dutta et al. (2021), Jiang et al. (2019), Pan et al. (2021), and Lindblad (2017), Golchin, and Rekabdar. (2024), Golchin and. Riahi (2021), among others.

The outcomes of these research efforts have led to the widely accepted perspective that economic uncertainty, policy uncertainty, and sentiment play significant roles in driving fluctuations in asset prices. By extension change in assets prices and market valuation directly influence the EV of firms. In the subsequent section, we offer a concise overview of some of the essential findings in this regard.

When uncertainty increases, it leads to a rise in risk premiums and interest rates, which in turn raises the cost of borrowing for most firms. This ultimately results in a decline in discounted cash flows and equity prices for affected firms. Consequently, both the market value and the enterprise value of the firm may suffer. Additionally, this negative relationship is likely to be more pronounced for firms that have higher financial leverage and fewer growth opportunities.

The substantial volume of research dedicated to investigating different aspects and characteristics of the airline industry reflects a profound interest in this sector of the economy. This emphasis on understanding the intricacies of the airline industry is warranted when considering its significant contribution to the overall economy. For instance, commercial air transportation played a substantial role, contributing approximately five percent, equivalent to $1.25 trillion, to the United States' Gross Domestic Product (GDP) in 2019.

Given the sizeable economic impact of the airline industry, it becomes evident that its vitality and the financial well-being of air carrier firms should not be underestimated. The multifaceted nature of the sector, encompassing various stakeholders, complex operations, and a diverse range of factors influencing performance, warrants a comprehensive understanding of its dynamics. Researchers, industry experts, and policymakers strive to delve into the intricacies of the airline industry to identify critical determinants of success, challenges faced by air carriers, and opportunities for growth and improvement.

By examining aspects such as safety practices, labor productivity, financial indicators, market share, and other relevant factors, researchers contribute to the development of knowledge that can guide decision-making processes and shape strategies within the airline industry. This deep interest in the sector points to its significance and the recognition that the financial health of air carrier firms can have far-reaching implications, not only for the industry itself but also for the broader economy. As such, the extensive literature on the airline industry serves as a valuable resource for stakeholders seeking to navigate the complexities of this vital sector and promote its sustainable growth and prosperity.

The goal of this research is to investigate the association of the air carrier firms' enterprise value (EV) with economic uncertainties. To this end, we construct an empirical model that considers relevant firm-specific financial and economic factors that may influence a carrier's EV in addition to market uncertainty indices.

The remainder of this paper is organized as follows. Section II presents a concise overview of the pertinent literature, highlighting key studies and research that inform the context of this paper. Data and their sources are described in Section III. Section IV delves into the methodology deployed in the paper. Section V presents an in-depth analysis and discussion of the empirical findings obtained through the applied methodology. The significance and implications of the results are examined, shedding light on their relevance to the research objectives and broader academic or practical contexts. Finally, the last section encapsulates the paper by providing a concise summary of the key findings and their implications. This section also includes the concluding remarks, discussing the broader implications, potential areas for further research, and any policy or practical implications that may arise from the findings.

## REVIEW OF THE RELEVANT LITERATURE

The sensitivity of the air transportation industry to market uncertainties is not a fixed attribute. Factors such as efficient cost management, strategic market positioning, and diversified service offerings can play a role in mitigating the impact of market uncertainties. Airlines with robust financial structures, flexibility in adapting to changing market conditions, and effective risk management practices are generally better positioned to navigate through economic turbulence.

Moreover, the industry's response to market uncertainties can also be influenced by external factors. Government policies, regulatory frameworks, and the overall health of the global economy can shape the magnitude of the impact. Collaborative efforts between airlines and industry stakeholders to address challenges and seize opportunities can further enhance the industry's resilience.

Numerous studies conducted by Bernanke (1983), Bloom (2009), Bloom et al. (2007) and Dixit (1989) have consistently demonstrated that uncertainty regarding the future plays a vital role in shaping economic behavior. One crucial factor contributing to this uncertainty is economic policy uncertainty (EPU), which stems from governmental decisions pertaining to fiscal, regulatory, or monetary matters. Governments hold a substantial stake in their respective economies, as evidenced by the significant proportion of federal, state, and local government expenditures in the United States, accounting for 42.45% of the gross domestic product in 2009. Consequently, any uncertainty surrounding the economic policies implemented by governments can have profound repercussions on financial markets, as highlighted by Bloom (2014).

According to Brogaard and Detzel (2015), economic policy uncertainty (EPU) has a profound impact on investment opportunities and leads to an increase in the equity risk premium. EPU, they argue, encompasses economic information that goes beyond general uncertainty, as it reflects the learning process of economic agents regarding the political costs associated with different policies. The shocks caused by news events that drive EPU may not carry the same level of risk pricing as those generating general economic uncertainty. If the shocks driven by news events are indeed priced, an effective measure of EPU can be used to predict expected returns while controlling for general uncertainty and economic distress.

In the context of the Intertemporal Capital Asset Pricing Model framework, it is expected that EPU would be associated with a negative price of risk. This implies that assets with lower exposure to EPU would serve as a hedge against the deterioration of investment opportunities and, consequently, have lower expected returns compared to assets with higher exposure. Therefore, constructing a portfolio that is long in assets with the lowest EPU exposure and short in assets with the highest EPU exposure should receive appropriate compensation for this risk differential.

Determining the sectors of the economy that exhibit greater resilience to market uncertainties is not a simple task. It is conceivable that sectors of the economy that face income elastic demand would also be more susceptible to the threat of economic uncertainty than those with inelastic demand, i.e., necessities. These sectors include healthcare, utilities, consumer staples, and telecommunications.

The resilience of these sectors stems from their ability to provide essential goods and services that are consistently in demand, regardless of the prevailing economic conditions. Healthcare remains a necessity regardless of economic fluctuations, as people require medical attention and treatments regardless of the state of the economy. Utilities, such as electricity and water services, are essential for everyday life, ensuring their demand remains relatively stable. Consumer staples encompass goods like food, beverages, and household products that people need on a day-to-day basis, making them less sensitive to economic fluctuations. Telecommunications, which encompass internet and communication services, are crucial for both personal and business needs, further enhancing their resilience.

It is worth noting that though these sectors are generally considered more resilient, they are not completely immune to market uncertainties. The impact of such uncertainties can still vary based on the specific circumstances. Factors like overall economic health, government policies, and industry-specific dynamics can influence how these sectors are affected. Therefore, while these sectors may exhibit greater resilience compared to others, careful analysis and monitoring are still necessary to fully understand their vulnerability to market uncertainties in any given situation.

Under typical economic circumstances, the air transportation industry is susceptible to market uncertainties due to its reliance on various factors. Elements such as fuel costs, passenger demand, geopolitical events, and economic trends collectively contribute to the industry's vulnerability. For instance, a decline in consumer confidence and disposable income can lead to reduced demand for air travel, negatively impacting the revenues and profitability of airlines. Furthermore, disruptions in global supply chains and trade can particularly affect cargo carriers within the air transportation industry.

While the air transportation industry is susceptible to market uncertainties, its ability to adapt, innovate, and respond to changing conditions can contribute to its overall resilience. Continuous monitoring, proactive management, and strategic decision-making remain crucial to mitigate risks and capitalize on potential growth opportunities in this dynamic sector.

The reaction of carriers' to changing market conditions such as fuel price uncertainties depend on income and price elasticity of demand for firms. For instance, with low-income elasticity, firms would be immune to economic booms and busts. Similarly, if the price elasticity of demand for the industry's services is low, at least in the short-run firms are not affected by sudden rise in fuel prices which may lead to fare hikes. The income and price elasticity of demand within the passenger air transportation industry can exhibit variations influenced by factors such as the region, market conditions, and the degree of competition. Air travel is commonly classified as income-elastic, signifying that an increase in income levels during prosperous economic conditions is associated with a greater rise in the demand for air travel.

Regarding price elasticity, demand for air travel is typically considered to be relatively inelastic, implying that changes in ticket prices have a relatively minor impact on demand. In other words, passengers are not highly responsive to price fluctuations when it comes to their decision to travel by air. However, it is important to note that during periods of economic recession or heightened competition, the price elasticity of demand for air travel may increase. This means that changes in ticket prices would have a more significant impact on the demand for air travel, leading to more pronounced shifts in consumer behavior.

The specific magnitude of income and price elasticity of demand within the passenger air transportation industry can vary across different regions and market conditions. Factors such as income distribution, cultural preferences, and availability of alternative transportation options can contribute to these variations. Additionally, the level of competition within the industry can also influence elasticity. In markets with limited competition, airlines may have more control over ticket prices and demand elasticity might be relatively low. Conversely, in highly competitive markets, airlines may need to adjust prices more actively, potentially increasing the price sensitivity of demand.

Beng, Ľ. F., & Hospodka, J. (2013), Fageda, X., & Flores-Fillol, R. (2012), Brons et al. (2002), Gallet, C. A., & Doucouliagos, H. (2014), Adrangi and Raffiee (2000), Peng et al. (2015), among others, have studied the price and income elasticities of demand for passenger air transportation.

These studies analyze the responsiveness of passenger demand to changes in ticket prices. They employ various econometric techniques and data sources to estimate these elasticities, considering factors such as income levels, market conditions, and other demand drivers. Some examples of these studies include Cigliano (1980) and Lo et al. (2015), among others.

Performance of an air carrier firm ultimately impacts its enterprise value. Numerous studies have delved into various factors that can influence the performance of air carrier firms. For example, Adrangi et al. (1997) conducted a study exploring the relationship between safety measures and the profitability of air carriers. Their research aimed to understand how safety practices and protocols impact the financial performance of these companies.

In a separate study, Adrangi et al. (1996) focused on investigating the influence of labor productivity on the profitability of air carrier firms. Their objective was to determine how the efficiency and productivity of labor within the industry affect the financial success of air carriers.

Furthermore, Adrangi and Hamilton (2023) conducted research that specifically examined the association of financial variables and market share with the performance of air carrier firms. They aimed to gain insights into how financial indicators and the market share of airlines contribute to their overall performance and competitiveness.

These studies collectively contribute to the body of knowledge surrounding air carrier performance by shedding light on different aspects of the industry. By exploring factors such as safety measures, labor productivity, financial variables, and market share, researchers aim to provide a comprehensive understanding of the determinants of success within the air carrier sector. Such insights can assist industry stakeholders in making informed decisions and implementing strategies to enhance the performance and profitability of air carrier firms under normal circumstances as well as in times of uncertainty and elevated risk. Furthermore, the factors associated with air carrier firms' performance and profitability may also impact their enterprise value.

In this section, we discuss the existing literature that addresses issues like those explored in this research, helping us narrow down the variables of importance for this work. The goal is to provide an overview of research on the relationship between economic uncertainties and the value or enterprise value of firms. Many existing studies do not directly delve into the determinants of EV. However, reviewing the extant literature provides clues regarding the relevant variables that may be associated with a firm's EV.

Brons et al. (2002) examine the sensitivity of passengers to prices as a means of predicting the efficacy of environmental policies in the aviation sector. The study focuses on analyzing the price elasticities of demand for passenger air transport to gain insights into price responsiveness within this specific context.

In conducting their research, the authors conducted a meta-analysis, compiling 37 studies with a total of 204 observations on price elasticities for passenger air transport. They systematically collected elasticity estimates and sample variables from each study, creating a comprehensive database for analysis. The price elasticity estimates identified fell within the range of approximately -0.3 to -1.7.

The results of the meta-regression analysis revealed that long-run price elasticities were higher than short-run elasticities, indicating that passengers tend to become more sensitive to price changes over time. Notably, business passengers exhibited lower price sensitivity, providing airlines with an opportunity to impose additional costs resulting from a price-based policy instrument without significantly impacting demand. The paper underscores the significance of considering these factors in the design of effective environmental policies within the aviation industry.

Adrangi et al. (2005) aimed to examine the relationship between real air transport equity index returns and expected and unexpected inflation, with the objective of determining whether investing in an air transport equity index serves as a reliable hedge against inflation. Previous research has indicated that increasing inflationary pressures have a negative impact on real economic activity, as well as the demand for equities and commodities. In line with these findings, this research supports a negative relationship between inflation and real air transport equity index returns, contradicting the Fisherian hypothesis for the specific index under investigation.

The findings of the study also reveal a significant and negative association between real air transport equity index returns and unexpected inflation, while no significant impact is observed with respect to expected inflation rates. This result suggests that unexpected inflation can influence the costs of fuel, labor, and meals, particularly due to the long-term contracts that air carriers often have with their labor unions. Additionally, renegotiations of contracts by pilots, flight attendants, and mechanics unions before their expiration can further contribute to the effects of unexpected inflation. Interestingly, the study suggests that airfares are not highly responsive to anticipated increases in general price levels. The negative relationship between real air transport equity index returns and the price level persists even after accounting for the negative effects of inflation on real economic activity. Consequently, unexpected inflationary pressures may adversely impact real air transport equity index returns. Hence, the study suggests that unexpected inflation negatively impacts equity returns and their market valuations, thus, the enterprise value of the firms in the sector.

Kang et al. (2014) investigate the impact of economic policy uncertainty and its components on firm-level investment. The sample data consists of an unbalanced panel of publicly traded manufacturing firms between 1985 and 2010. The annual firm-level investment data come from Standard and Poor's Industrial Annual COMPUSTAT database, and the monthly economic policy uncertainty index comes from Baker et al. (2013). The results indicate that economic policy uncertainty in combination with firm-level uncertainty reduces firms' investment decisions, particularly for firms with higher firm-level uncertainty and during a recession. The study also finds that policy uncertainty does not seem to affect the investment decisions of the largest firms. The economic policy uncertainty index is a weighted average of four uncertainty components: news-based policy uncertainty, CPI forecast interquartile range, tax legislation

Brogaard and Detzel (2015) test the hypothesis that economic policy uncertainty increases equity risk premium (ERP) and is priced in the cross section of stock returns. They examine the impact of EPU on asset prices in both the time series and cross section. They highlight the importance of government economic policy and the difficulty of diversifying against it. The authors use a measure of policy uncertainty developed by Baker et al. (2013), which is a weighted average of three components - the frequency of major news discussing economic policy-related uncertainty, expiring tax provisions, and forecaster disagreement about government purchases and inflation. The Baker et al. (2013) measure is particularly appealing because it provides continuous tracking of policy risk compared to other alternatives (Baker et al., 2014). Research has shown that EPU affects investment opportunities and increases the equity risk premium. Some argue that EPU contains economic information beyond general uncertainty, as it reflects agents' learning about the political costs of different policies. News shocks that drive EPU may not carry the same price of risk as those driving general economic uncertainty. If news-based shocks are priced, a good measure of EPU can help predict expected returns while controlling for general uncertainty and economic distress.

Their study uses Merton's (1973) capital asset pricing model (CAPM) framework and causal mechanisms from Pastor and Veronesi's (2012,2013) models to analyze the relationship between risk and return. The results suggest that both general economic uncertainty and EPU increase during periods of investment opportunities. They find that EPU helps forecast log excess returns on the stock market beyond other measures of uncertainty, and innovations in EPU command a significant negative risk premium in the cross section of stock returns, even when controlling for other factors. Thus, investors are compensated with higher expected returns for bearing systematic market risk and the risk of unfavorable shifts in the investment opportunity set. Additionally, assets that hedge against adverse changes in economic uncertainty have greater demand, bidding up their prices and driving down their expected returns.

Brogaard and Detzel (2015) offer the following explanation. In the CAPM framework, EPU should carry a negative price of risk, meaning assets with lower exposure to EPU would act as a hedge against investment opportunity deterioration and have lower expected returns than those with greater exposure. A portfolio long in assets with the lowest EPU exposure and short in assets with the highest EPU exposure should be compensated accordingly.

Borghesi and Chang (2020) investigate the correlation between global economic policy uncertainty, intangible assets, research and development (R&D), and firm valuation. The data analyzed spans the years 1997 to 2015 and encompasses various variables sourced from diverse outlets, including Worldscope for firm financials, the World Bank for country-level macroeconomic factors, and ASSET4 for scores related to corporate governance, employee policies, environmental practices, and social responsibility. To assess the impact of economic instability, the authors employ the global economic policy uncertainty index by Baker et al. (2016), specifically focusing on 40 countries. This index is derived as the annual average of monthly GDP-weighted global economic policy uncertainty (GEPU), normalized to a mean value of 100. Tobin's q is utilized as a metric for firm value.

Through multiple regression analyses, Borghesi and Chang identify that companies operating in industries with high intangible intensity and those involved in R&D are most adversely affected by stringent governance policies in times of elevated economic instability. Conversely, during periods of economic volatility, these same companies experience the greatest benefits from previous investments in corporate social responsibility (CSR). In summary, the study concludes that past CSR investments serve to preserve the value of firms with a high intensity of intangible assets during times of heightened economic policy uncertainty, while strict corporate governance mechanisms diminish the value of such firms.

Duong et al. (2020) address the escalating concerns regarding the influence of economic policy uncertainty (EPU) on various corporate activities, including capital investment, mergers and acquisitions, innovation, corporate transparency, as well as equity prices and risk premia. Their study investigates whether corporate managers alter their cash holdings strategy in response to heightened policy uncertainty and, if so, whether and how these adjustments in cash holdings contribute to alleviating the adverse impact of policy uncertainty on the tangible economic activities of firms. The authors specifically concentrate on the impact of macro-level EPU on firm cash holdings, recognizing the pivotal role of cash holdings in influencing other corporate financial policies. They argue that corporate managers are more inclined to be responsive to EPU when managing cash holdings compared to making long-term-oriented capital structure decisions. To scrutinize the relationship between policy uncertainty and corporate cash holdings, they utilize the economic policy uncertainty index by Baker et al. (2016) in conjunction with the Compustat annual industrial file, covering the period from 1985 to 2014.

Their regression findings reveal a robust positive correlation between economic policy uncertainty and corporate cash holdings among US firms spanning the period from 1985 to 2014. This relationship persists even when accounting for firm investment opportunities and macroeconomic uncertainty. The consistency of the results holds across various measures of policy uncertainty and is particularly notable for companies characterized by high government dependence, exposure to political risk, and sensitivity of returns to changes in policy uncertainty. The study also indicates that, in times of increased policy uncertainty, firms tend to accumulate more cash, and managers opt for cautious payout decisions. The heightened cash holdings serve to mitigate the adverse impact of policy uncertainty on both capital investment and the innovation output of firms. Importantly, the observed link between policy uncertainty and cash holdings is not attributable to delays in investment but rather stems from the aggravation of external financing conditions during periods of uncertainty. This research enhances our understanding of the effects of policy uncertainty on corporate financial constraints and behaviors, contributing valuable insights to the literature on the role of corporate cash holdings in alleviating the negative consequences of policy uncertainty on the tangible economic activities of firms.

Jory et al. (2020) examine the impact of economic policy uncertainty (EPU) on firms' trade credit policy and its effect on firm value. The study uses a large sample of 1,472,930 firm-quarter observations from the Compustat Fundamentals Quarterly database from 1980 to 2017.

Their study investigates the impact of government economic policy uncertainty on trade credit and firm value. It finds that during periods of high economic policy uncertainty (EPU), firms reduce their receivable days and face lower payable days from their suppliers. The negative effect of EPU on trade credit is consistent across various firms. Tightening credit extensions is positively related to firm value, but an overly conservative trade credit policy implemented during high EPU periods may drive customers to competitors. The study has important implications for corporate managers and policymakers. The findings could be relevant for U.S. public firms but do not necessarily extend to private ones. Additionally, the study examines the impact of EPU and trade credit policy on firm value for the subsamples of firms in industries with high vs. low concentration.

Zhu et al. (2020) explore the impact of economic policy uncertainty (EPU) on enterprise value, utilizing China's EPU index and financial data from A-share listed enterprises spanning the period 2004 to 2018. Their empirical findings reveal a notable inhibitory effect of economic policy uncertainty on enterprise value overall, with the magnitude of this impact closely tied to various enterprise characteristics. These characteristics include financial leverage, scale, research and development intensity, the level of marketization, ownership structure, and geographical location. Specifically, an increase in financial leverage and scale tends to alleviate the constraining effect of EPU on enterprise value. Conversely, an increase in R&D intensity and the degree of marketization exacerbates this effect. Furthermore, non-state-owned enterprises (non-SOEs) and firms situated in first-tier cities are identified as more susceptible to the adverse impacts of escalating EPU. Their findings underscore the importance for policymakers to prioritize macroeconomic policy stability, and for micro-enterprises to proactively consider and prepare for the repercussions of economic policy uncertainty.

El Ghoul et al. (2023) examine the impact of EPU on the cash holdings of multinational corporations using a substantial dataset. Elevated EPU raises concerns among investors about managerial self-dealing and political extraction. As a result, the perceived potential cost of holding cash (associated with expropriation) outweighs its benefits (such as precautionary motives), leading to a reduction in the optimal level of cash holdings. Their findings provide supporting evidence that firms tend to decrease their cash reserves when faced with high EPU. Additionally, they demonstrate that the market devalues excess cash holdings during periods of heightened policy uncertainty. However, this negative effect is moderated by factors such as robust investor protection, a more independent press, and higher government quality.

Jumah et al. (2023) utilize firm-level data spanning twenty-two countries from 2000 to 2020, encompassing 364,433 firm-year observations of 29,709 distinct firms. Index-based measures are employed for both EPU and corporate diversification. Their findings reveal a negative impact of EPU on firm value. However, corporate diversification effectively moderates this adverse impact by adeptly alleviating the constraints imposed by financial uncertainties. Further analysis indicates that both related and unrelated diversification play a crucial role in mitigating the negative effects of high EPU on firm value in developed economies. In contrast, in emerging economies, only unrelated diversification proves effective in addressing elevated EPU. The robustness of their results is affirmed through subsampling, sensitivity analyses, and addressing endogeneity concerns. This study implies a strategic recommendation for managers – diversification serves to sustain firm value amidst uncertainty. Additionally, policymakers should perceive heightened policy uncertainty as a threat to the stability of the business environment and take measures to minimize uncertainty, fostering a more conducive environment for businesses.

Other scholars have explored the association of economic policy uncertainty and mergers and acquisitions (Bonaime et al. (2018)), uncertainty and economic cycles (Ludvigson (2021), economic policy uncertainty in China (Davis et. al. (2019).

It is clear from the literature that economic uncertainties have an impact on the enterprise value of a firm. Enterprise value is a measure of a company's total value, including its market capitalization, debt, and cash. Economic uncertainties, such as changes in interest rates, inflation, and GDP growth, can affect a company's expected future cash flows and risk profile, which can in turn impact its enterprise value. For example, if economic uncertainty increases, investors may require a higher rate of return to compensate for the increased risk, which could reduce the present value of the company's future cash flows and lower its enterprise value. The present research aims to investigate the association of the EV of a select group of the US air carrier firms and uncertainties esteeming from global and domestic sources Thus, this study contributes to the existing literature by delving into air travel sector. In the next section we briefly explain the data and the methodology of our research.

## DATA AND METHODOLOGY

The data for the study are derived from Compustat database of WRDS. The period of the study spans from the first quarter of 1980 to the fourth quarter of 2022. The data for some carriers are not available for the entire sample period. In those cases, we extract the available data. We chose five major air carrier firms, Alaska Airlines (ALA), American Airlines (AA), Delta Airlines, Southwest Airlines (SWA), United Airlines, and Skywest Airlines (SKA). The first five airlines are considered major carriers with significant market shares. SKA is a smaller player with mainly domestic routes. We include this firm as a representative for smaller firms, with mainly domestic operations, and limited market share.

Figure 1 shows the share of US domestic market share by carriers from February to January 2023. American Airlines, Delta Airlines, Southwest Airlines, United Airlines, Alaska Airlines, and Skywest Airlines and each accounted for 17.5, 17.3, 16.9, 15.6, 6.2 and 2.8 percent of the sales in the same period.

Figure 2 shows the share of US domestic market share of the air carriers based on the passenger count from January to October 2018. American Airlines, Delta Airlines, Southwest Airlines, United Airlines, Alaska Airlines, and Skywest Airlines and each accounted for 15, 16, 20, 11, 4 and 6 percent of the total domestic passengers carried during this period.

Figure 3 shows the US domestic market capitalization of the major world air carriers, American Airlines, Delta Airlines, Southwest Airlines, and United Airlines are the top-ranking world companies based on market capitalization in 2022. It is clear from figures 1 through 3 that Delta, American and United airlines are ranked among the top air carriers in the world by all commonly used criteria.

Regarding its network size, American Airlines surpasses all other US and global carriers. The airline provides flights to approximately 364 destinations globally, including 126 international destinations across 64 countries. This extensive network constitutes around 34.6% of American Airline's complete operations.

United Airlines boasts the second-largest route network within the United States. However, it outperforms American Airlines in terms of the number of countries it serves. As reported by Canada.com, United Airlines operates across 65 countries worldwide.

Like American Airlines, the majority of United's flights are concentrated within domestic destinations. The airline's routes cover 235 cities within the United States, accounting for approximately 64.9% of its overall route network. The remaining routes are allocated to international flights, and once again, United is actively pursuing expansion. The airline announced its intentions to initiate new routes, including connecting San Francisco International Airport and Dublin Airport.

Delta Air Lines, positioned as one of the "big three" airlines, holds the distinction of having the third most extensive route network among US carriers. The airline's operations encompass 326 destinations, spanning 57 countries and featuring 99 international routes. A substantial portion, approximately 69.6%, of Delta's route network is dedicated to domestic flights. Notably, the airline has unveiled plans to introduce new services including connecting Boston, Massachusetts, with Rome, Italy.

Alaska Airlines operates primarily within the domestic sphere, servicing a portfolio of over 100 destinations. The airline's daily operations comprise approximately 1,200 flights, spanning the United States, Mexico, Canada, and Costa Rica. In total, Alaska Airlines provides access to a network of 115 routes.

Southwest Airlines presents a network of 103 routes spanning the United States, Mexico, the Caribbean, and Central America. Among its notable and frequently traveled routes are connections between Dallas and Los Angeles and Atlanta to Chicago, among others.

Among U.S. airlines, higher-margin international travel is the most important to United, accounting for about 38% of its passenger revenue before the pandemic. In the second quarter, international passenger revenue accounted for about 41% of the airline's total passenger revenue.

Delta Airline’s largest market share is in the International Airlines industry, where they account for an estimated 77.6% of total industry revenue and are considered an All Star because they display stronger market share, profit and revenue growth compared to their peers.

To observe the reaction of the enterprise value of each firm to shocks in financial and uncertainty indices, we estimate a VAR model for each carrier. This method reveals that the effects of shocks to EPU, GEPU, VIX, and recessionary risk differ across firms with varying financial structures.

The quarterly financial variables that may impact the EV are as follows:

Current ratio (CR)

Debt to asset ratio (DA)

Market share (MKSHARE) is the square of the ratio of a carrier's revenues to the total revenue as in Herfindahl index

GDP growth rate (GDPGRW)

Operating income after debt (OIAD)

Enterprise value (EV)

Averages of these variables are presented in Table 2.

The monthly variables that account for economic policy and market uncertainties that potentially impact the EV are as follows.

Economic policy uncertainty index (EPU)

Global economic policy uncertainty index (GEPU)

Recession Risk (RECRISK)

VIX

The first two are provided by Baker et al (2016). Recession risk is the smoothed U.S. recession probabilities, not Seasonally Adjusted. These series are taken from taken from the Federal Reserve Bank of St. Louis database or FRED. The monthly VIX values are from Yahoo Finance.

The mainstay of our methodology is vector autoregressive model (VAR). The VARs are particularly adept at modeling systems of interrelated variables. VARs are extensively described and utilized in the literature. In this section, we offer a summary of this methodology. The reduced form VAR that we deploy avoids the complex structure of modeling the interactions of variables by treating all model variables as endogenous. This approach is superior to others that require pre-determined role as exogenous and endogenous variables because the interconnection among economic and financial variables is not completely cut and dry.

A VAR model of order p may be expresses as equation (1)

$$y_t = A_1 y_{t-1} + A_2 y_{t-2} + .... + \varepsilon_t \ , \qquad (1)$$

where,

$y_t = (y_{1t} y_{2t} .....y_{jt})'$ is a vector of Jx1 model endogenous variables,

$A_1 A_2 .....A_p$ is the matrix of model coefficients that will be estimated.

$\varepsilon_t = (\varepsilon_{1t} \varepsilon_{2t} ....\varepsilon_{jt})$ is a j by1 vector of white noise model innovations with the following matrix of variances and characteristics.

$E(\varepsilon_t) = 0, E\ (\varepsilon_t \varepsilon_t') = \Sigma_\varepsilon$

$E(\varepsilon_t \varepsilon_s') = 0$ for $t \neq s.$

We deploy the Mixed Frequency Data Vector Autoregressive (MIDAS-VAR) model which is a powerful variation of the standard VARs that are used to analyze relationships between variables with different sampling frequencies. This approach is particularly useful when dealing with economic data that is observed at different frequencies, such as monthly, quarterly, or annual data. MIDAS-VAR models allow for the integration of high-frequency and low-frequency data, enabling researchers to capture the dynamics and relationships between variables at different time scales.

The basic idea behind MIDAS-VAR models is to use lagged values of high-frequency data to explain the current or future values of low-frequency data. This is achieved by estimating a dynamic relationship between the variables using a combination of high-frequency and low-frequency data.

We will offer a brief explanation of a simple MIDAS-VAR with m high frequency periods per low frequency observation. For instance, for each monthly observation on a variable, there may be twenty daily observations on another set of variables. To demonstrate a MIDAS -VAR model, let us assume that a VAR consists of $n_{Lt}$ variables observed at low frequency and $n_{Ht}$ , at high frequency. Forming a VAR from the high and low frequency variables in a VAR produces the system of equation (2)

$$\begin{bmatrix} Y_{H.t_t,1} \\ Y_{H.t_t,2} \\ \cdots\cdots \\ Y_{H.t_t,n} \\ Y_H, t_L \end{bmatrix} = \begin{bmatrix} \Gamma_j^{1,1} & \Gamma_j^{1,2} & \cdots & \Gamma_j^{1,n} & \Gamma_j^{1,n+1} \\ \Gamma_j^{2,1} & \cdots & \cdots & \cdots & \Gamma_j^{2,n+1} \\ \cdots & & & & \cdots \\ \Gamma_j^{n,1} & \cdots & \cdots & \Gamma_j^{n,n} & \Gamma_j^{n,n+1} \\ \Gamma_j^{n+1,1} & \cdots & \cdots & \Gamma_j^{n+1,n} & \Gamma_j^{n+1,n+1} \end{bmatrix} \begin{bmatrix} Y_{H.t_t-j,1} \\ Y_{H.t_t-j,2} \\ \cdots \\ Y_{H.t_t-j,n} \\ Y_{L.t_t-j} \end{bmatrix} + \begin{bmatrix} E_{H.t_L,1} \\ E_{H.t_L,2} \\ \cdots \\ E_{H.t_L,n} \\ E_H, t_L \end{bmatrix}, \quad (2)$$

where matrix $\Gamma_j^{a,b}$ is $K_H*K_H$, $\Gamma_j^{n+1,b}$ $K_L*K_H$, $\Gamma_j^{a,n+1}$ $K_H*K_L$ for all j, a,b= 1,2,.. n, and $\Gamma_j^{n+1,n+1}$ is $K_L*K_L$.

In this model, $Y_{Ht}$ and $Y_{Lt}$ represent the high and low-frequency variables observed at time t. The lagged values of Y, denoted as Y(t-j), capture the autoregressive component of the low-frequency variable. The coefficient matrices $\Gamma_j^{\alpha,\beta}$ represent the parameters to be estimated, and $E_{H,t_L}$ and $E_{L,t_L}$ are the high and low frequency white noise error term vectors.

Ghysels et al. (2004 and 2016) offers the Bayesian MIDAS-VAR estimation. This methodology requires specifying prior distributions for the model coefficients and the covariance matrix of the white noise residuals.

Combining the specified priors with VAR likelihood generates the conditional posteriors of the model coefficients as well as the innovations covariance matrix as

$(\beta|y,\Sigma) \sim N(\bar{\beta},\bar{\Omega})$ and

$(\Sigma|y,\beta) \sim IW(\bar{Q},\bar{\varpi})$,

where,

$$\bar{\Omega} = (\underline{\Omega}^{-1} + (\hat{\Sigma}^{-1} \otimes (X''X')^{-1}))^{-1}$$

$$\bar{\beta} = \bar{\Omega}(\underline{\Omega}^{-1}\underline{\beta} + (\hat{\Sigma}^{-1} \otimes (X'')y)$$

$$\bar{Q} = \underline{Q} + E'E$$

$$\varpi = \underline{\varpi} + T$$

The model parameters' prior distribution adheres to Litterman's (1986) approach, where the prior means are set to zero except for the first lag terms. Ghysels et, al (2016) introduces modifications to Litterman's assumptions by assigning a distinct hyper-parameter to each own lag term based on its observation frequency. Additionally, the high-frequency series variables are assumed to follow an AR (1) process, leading to exponential parameters in the low-frequency domain. The prior variance of the model coefficients also incorporates cross-frequency variances.

MIDAS-VAR models offer several advantages in econometric modeling. They allow for the incorporation of high-frequency data, which captures more detailed information and short-term dynamics, into the analysis of low-frequency variables. This enables researchers to better understand the relationships and transmission mechanisms between economic variables at different time scales.

Furthermore, MIDAS-VAR models can improve forecasting accuracy by exploiting the additional information contained in high-frequency data. By incorporating the dynamics of both low-frequency and high-frequency variables, these models provide a more comprehensive and accurate picture of the economic system under study.

In summary, MIDAS-VAR models are a valuable tool in econometric modeling, particularly when dealing with mixed frequency data. They allow for the integration of variables observed at different frequencies and provide insights into the dynamic relationships between them. By capturing the short-term and long-term dynamics of economic variables, MIDAS-VAR models enhance our understanding of complex economic systems.

To ensure that the model variables are stationary, we deploy two standard tests, the Augmented Dickey-Fuller and Phillips-Perron.

## EMPIRICAL FINDINGS

Table 2, panel A reports the means of financial variables for the five air carries under study. It shows that the air carrier firm under study vary with respect to CR, DA, Market Share and OIAD. Therefore, it is reasonable to expect variations in the response of EV to each of these financial factors.

The panel B of Table 2 presents the findings of the Augmented Dickey-Fuller (1979) and Phillips-Perron (1987) tests of stationarity. VAR-MIDAS modeling, like most VARs require stationary model variables. In cases that model variables of risk indices are not stationary, we induce stationarity by first differencing the nonstationary variables.

The reaction of the EV of each air carrier to shocks to economic, financial, and risk indicators appear to depend on each firm's business model, market positioning, as well as its management's ability to react. Therefore, we expect differences in the reaction of EV to uncertainty across firms. The VAR-MIDAS models estimated for each airline produce impulse responses (IR) like any VAR model. An initial review of IRs indicate that they behave erratically with no clear direction beyond four quarters. It is plausible that the impact of most shocks to EV would not last beyond four quarters. Therefore, the focus of our attention in on the four quarters after one standard deviation shocks to the model variables.

Examining Figures 4 through 9, the accumulated effect of one standard deviation positive shock to every economic and economic policy uncertainty index except for VIX on EV of every firm is negative or mixed. Therefore, accounting for several relevant financial variables, the EV of most air carrier firms is sensitive to the risk of recession, domestic economic policy uncertainty, and GEPU. The negative reaction of the EV of US air carriers to various domestic and global uncertainty indices corroborate the theoretical expectations.

Turning to the accumulated impulse response of EV to the financial variables in the model, namely CR, DA, OIAD, and MKS, Figures 4 through 9 show that the accumulated response of EV of all air carriers to shocks to the current ratio are positive for all firms except Delta and Skywest Airlines which show a mixed accumulated reaction. Similarly, the accumulated impulse response of EV of all firms to shocks to the debt to asset ratio are generally negative again except for Delta airlines. This airline has a significant international presence which maybe shielding it from domestic recessionary threats. In most cases, shocks to the OIAD results in a positive accumulated effect on the impulse responses of the EV, as expected. Shocks to the remaining financial variables result in a mixed accumulated EV response of the air carrier firms.

Table 3 summarizes the findings of accumulated impulse responses of the EV to various financial variables included in the model. While there are no unanimously consistent responses to shocks to financial variables, CR, DA, and OIAD stand out. For majority of the airlines under consideration the impulse response of the EV to shocks to CR and OIAD are positive. The reaction of the EV to positive shocks to DA is negative indicating that the higher debt to asset ratio is associated with falling EV. In the remaining cases there is no clear reaction by EV to positive shocks to remaining financial variables. These findings consistent with our expectations. In Table 3, it is evident that most air carriers in the sample exhibit a Debt-to-Asset higher than the average of all US companies in 2021, as per publicly available data. Consequently, a higher DA may be perceived unfavorably by analysts and the market. Additionally, Table 2 affirms that the market shares of the air carrier under study are relatively similar and may not be a crucial variable in association with variations in EV.

A higher current ratio indicates that the company has more current assets relative to its current liabilities. This suggests that the company is better positioned to meet its short-term obligations, including paying off debts and covering operational expenses.

Lenders and investors often view companies with a higher current ratio more favorably because it implies a reduced risk of default on short-term obligations. This can lead to lower perceived financial risk, which can positively impact the company's cost of capital.

A rise in the current ratio may also be an indicator of efficient working capital management. Companies with effective inventory and receivables management can maintain higher current ratios without excessive amounts of idle or non-performing assets, which can contribute to higher overall efficiency.

A healthy current ratio provides a company with strategic flexibility. It means the company has the resources to take advantage of unexpected opportunities, weather economic downturns, or invest in projects that contribute to long-term value creation.

Investors often look for signs of financial stability and prudence. A higher current ratio can instill confidence in investors, as it suggests that the company is managing its short-term obligations responsibly and is less likely to face financial distress.: Enterprise value is calculated as the sum of a company's market capitalization, debt, minority interest, and preferred shares, minus its cash and cash equivalents. As the current ratio improves, it positively influences the financial metrics considered in enterprise valuation, potentially leading to a higher enterprise value.

Similarly, rising operating income after depreciation can positively impact a firm's enterprise value for several reasons. Operating income after depreciation is a key indicator of a company's profitability from its core operations. As this figure rises, it suggests that the company is generating more income from its day-to-day business activities, which is a positive signal for investors and stakeholders.

Operating income after depreciation is also a measure of cash flow generated by a company's core operations. The depreciation component is a non-cash expense, meaning it doesn't involve an actual cash outflow. Therefore, when operating income increases after accounting for depreciation, it reflects an improvement in the company's ability to generate actual cash. Higher operating income after depreciation improves a company's ability to service its debt obligations. Lenders often look at a company's cash flow to assess its capacity to meet interest and principal payments. A firm with increasing operating income after depreciation is better positioned to cover its debt service requirements. Furthermore, a growing operating income provides a company with more resources to reinvest in its business. Whether it's expanding operations, investing in new projects, or conducting research and development, an increase in operating income after depreciation indicates improved capacity for internal investment.

A firm with rising operating income after depreciation is generally considered less risky. It suggests that the company's core business operations are becoming more profitable. On the contrary a rising debt-to-asset ratio can negatively impact a firm's enterprise value for several reasons as follows.

The accumulated impulse response of EV to DA is negative for most air carriers. A higher debt-to-asset ratio indicates that a larger portion of the firm's assets is financed through debt. While debt can be a useful tool for financing operations and expansion, a high level of debt also increases financial risk. Investors may become concerned about the company's ability to meet its debt obligations, especially if economic conditions or the company's financial performance deteriorate. Airline companies tend to operate at high debt to equity ratios. For instance, Southwest Airlines shows the lowest debt to equity ratio of 90 percent while American Airlines operates with the highest debt to equity ratio of 446 with other companies in the sample somewhere in between these two ratios.  Consumer of energy firms often operate at much lower debt to equity ratios.  For example, Exxon-Mobile and Proctor and Gamble operate with debt-to-equity ratios of 63 and 19 percent, respectively.

As the debt-to-asset ratio increases, lenders and investors may perceive the company as riskier. To compensate for the increased risk, lenders may demand higher interest rates on new debt, and investors may require a higher expected return on the company's equity. This can result in a higher overall cost of capital for the firm, reducing its valuation.

A rising debt-to-asset ratio can also limit a company's financial flexibility. High levels of debt mean that a significant portion of the company's cash flow may be allocated to servicing debt payments, leaving less room for strategic investments, research and development, and other value-enhancing activities. This limited financial flexibility can constrain the company's ability to adapt to changing market conditions.

As the debt-to-asset ratio increases, credit rating agencies may downgrade the company's credit rating. A lower credit rating can further raise the cost of borrowing and negatively affect the firm's reputation among investors. This downgrade can have a cascading effect on the company's enterprise value.

Investors and analysts often view a high debt-to-asset ratio as a sign of financial distress or instability. This negative perception can lead to a lower valuation of the company in the stock market. Investors may sell their shares, causing a decline in the stock price and, consequently, the firm's enterprise value. During economic downturns or periods of financial stress, companies with high debt levels may face increased challenges in meeting their debt obligations. This vulnerability to economic downturns can lead to a further decline in the company's enterprise value.

In summary, an increase in the current ratio or operating income after depreciation (OIAD) contributes to a company's financial health, reduces risk, and enhances its overall attractiveness to investors and lenders, all of which can positively impact its enterprise value. However, a high debt-to-asset ratio could negatively affect the firm's enterprise value, especially if firms are heavily burdened with debt, as is often the case with most air carriers. Additionally, positive shocks to market share may elicit mixed reactions in the enterprise value of air carriers.

Figure 4 presents the summary of accumulated impulse responses of the EV to shocks to uncertainty indices. The enterprise value of a company may react negatively to risks of recession and policy uncertainties due to several interconnected factors that impact investor sentiment, financial performance, and overall market conditions. There are key reasons why these factors can adversely affect enterprise value.

During a recession, consumer spending tends to decline, and businesses may experience lower demand for their products and services. This can lead to a decrease in the company's revenue and profitability. Investors often value companies based on their future expected cash flows, and a recession can raise concerns about the firm's ability to maintain or grow its earnings, negatively affecting its enterprise value.

Economic downturns normally increase the risk of default on debt obligations for companies, especially those with high levels of debt. As recessionary pressures mount, companies may struggle to meet their interest and principal payments. This heightened default risk can result in a decrease in the company's credit rating, leading to higher borrowing costs and a potential decline in enterprise value.

Fears of recession and Economic policy uncertainties can contribute to increased market volatility. Investors may become more risk-averse, leading to higher market discount rates. Elevated volatility can negatively impact stock prices and valuation multiples, affecting the enterprise value of companies across various sectors.

During a recession, central banks may implement monetary policies such as lowering interest rates to stimulate economic activity. While this can reduce borrowing costs for companies, it may also lead to concerns about the overall economic health. Investors may interpret interest rate cuts as a response to economic challenges, contributing to negative sentiment and potentially impacting enterprise values.

The EV of many of the airlines under study is sensitive to VIX, the risk of recession as well as economic policy uncertainty. The reaction of the EV to the GEPU and EPU for most carriers in the sample is negative.

Uncertainty regarding government policies, such as changes in tax regulations, trade policies, or industry-specific regulations, can create challenges for companies in planning and decision-making. Investors may respond to policy uncertainties by discounting the future cash flows of a company more heavily, resulting in a lower valuation and enterprise value.

Economic uncertainties and recessionary fears can erode consumer and investor confidence. Reduced confidence may result in decreased spending, investment, and overall economic activity. This lack of confidence can lead to a sell-off in the stock market, affecting equity values and, consequently, the enterprise values of companies. The demand for air travel tends to be income elastic and sensitive to economic downturns and policy uncertainties. Investors may reevaluate their expectations for companies in this industry, leading to lower enterprise valuations.

The accumulated impulse response of the EV of most air carriers to GEPU is negative as expected. Global economic uncertainty can negatively impact the enterprise value of a company due to several interconnected factors. Firstly, uncertainty on a global scale often leads to market volatility, which can result in fluctuations in stock prices and overall market instability. This volatility makes it challenging for companies to accurately forecast their future earnings and plan their investment strategies, thereby reducing investor confidence and potentially lowering the company's valuation.

Secondly, global economic uncertainty can disrupt international trade and supply chains, affecting companies that rely on global markets for their revenue streams or source materials for their production processes. Trade tensions, geopolitical conflicts, or unexpected economic downturns in key trading partners can lead to decreased demand for goods and services, reduced sales, and ultimately lower profitability for affected companies.

Additionally, global economic uncertainty may impact currency exchange rates, interest rates, and inflation levels, all of which can have significant implications for businesses operating in international markets. Fluctuations in exchange rates can affect the competitiveness of exports and imports, while changes in interest rates and inflation can impact borrowing costs and consumer purchasing power, respectively, further affecting companies' financial performance and enterprise value.

Moreover, uncertainty about the future economic landscape may lead to cautious consumer and investor behavior, resulting in reduced spending, delayed investment decisions, and lower revenue generation for companies across various sectors. This decrease in economic activity can have a cascading effect on corporate earnings and ultimately impact the valuation of companies in the form of lower stock prices or reduced acquisition premiums.

The response of the EV of most air carriers to shocks to VIX is positive. Positive shocks to the VIX, despite being associated with increased market volatility, can paradoxically raise the enterprise value of air carrier firms due to several key dynamics. Firstly, heightened volatility in the broader market often leads to a flight to safety among investors, with traditional safe-haven assets such as bonds becoming more attractive. However, air carrier firms, being non-cyclical and essential providers of transportation services, can be perceived as relatively resilient in turbulent market conditions. Consequently, investors seeking refuge from market uncertainty may view air carrier stocks as comparatively stable investments, leading to increased demand for these shares and a subsequent rise in their market value. Moreover, in times of elevated volatility, air carrier firms may have the opportunity to capitalize on market dislocations and implement strategic initiatives such as route optimization, capacity adjustments, or cost-cutting measures, which can enhance operational efficiency and profitability. These factors collectively contribute to a positive impact on the enterprise value of air carrier firms despite the presence of heightened volatility signaled by a positive shock to the VIX.

In summary, a rising debt-to-asset ratio can hurt a firm's enterprise value by increasing financial risk, raising the cost of capital, limiting financial flexibility, impacting credit ratings, influencing market perception, and exposing the company to economic vulnerabilities. The enterprise value of a company may react negatively to risks of recession and policy uncertainties due to concerns about revenue and profitability, increased default risk, market volatility, policy uncertainties, interest rate movements, confidence levels, and industry-specific risks. Investors often reassess their valuation models in response to changing economic conditions and uncertainties, which can impact the perceived value of a company. The global economic uncertainty introduces a range of risks and challenges for businesses, including market volatility, disruptions to trade and supply chains, currency fluctuations, and changes in consumer and investor behavior. These factors can collectively contribute to a negative impact on the enterprise value of a company by affecting its financial performance, growth prospects, and investor confidence in uncertain times. In specific economic climates, positive shocks to the VIX may elicit positive reactions from the VIX, but this isn't always the case.

## SUMMARY AND CONCLUSIONS

Enterprise value holds significance in company valuation as it considers not only the company's equity but also all its assets and liabilities. This renders it a more precise indicator of the company's total value, particularly when assessing companies with varying capital compositions.

The connection between economic uncertainty and firm value finds its roots in economic theory. Early explorations of how economic uncertainty influences corporate behavior can be traced back to Sandmo's pioneering work in 1971. Following John Kenneth Galbraith's "The Age of Uncertainty" in 1977, numerous significant events, extensively covered by both the media and academia, have highlighted the central role of uncertainty in the financial and economic landscape.

Expanding on this foundational research, subsequent scholars, such as Flacco and Kroetch ( 1986), Fooladi and Kayhani (1990 and 1991), and Adrangi and Raffiee (1990), have delved into various aspects of how firms respond to uncertainty. These researchers have embarked on comprehensive theoretical investigations to uncover the intricate ways in which companies may adapt their production methods, pricing strategies, and profit-seeking endeavors when confronted with market risks and uncertainties.

In recent times, the research conducted by Baker et al. (2016) has been instrumental in quantifying and examining economic and economic policy uncertainties. Their efforts have involved the creation of indices, offering valuable perspectives into the ever-evolving landscape of economic uncertainty. These insights have proven invaluable for policymakers, economists, and investors, enhancing their comprehension of these vital aspects.

In normal economic conditions, the aviation sector is exposed to market uncertainties owing to its dependence on a multitude of factors. Factors like fuel expenses, passenger demand, geopolitical incidents, and economic patterns collectively contribute to the sector's susceptibility. For instance, a decrease in consumer confidence and disposable income can result in decreased air travel demand, which can adversely affect the earnings and profitability of airlines. Additionally, disturbances in global supply networks and trade can have a notable impact, particularly on cargo carriers within the aviation industry.

However, it's crucial to highlight that the air transportation industry's susceptibility to market uncertainties is not a static characteristic. It varies based on specific conditions and situations, and some airlines may demonstrate greater resilience in navigating economic downturns compared to others. As an illustration, in 2019, commercial air transportation made a significant contribution, accounting for approximately five percent of the United States' Gross Domestic Product, which equated to a staggering $1.25 trillion.

The aim of this study is to explore the association between the enterprise value of air carrier firms and economic uncertainties. To achieve this objective, we develop an empirical model that takes into consideration pertinent company-specific financial and economic factors that may exert an influence on a carrier's EV, in addition to market uncertainty indices.

The accumulated impulse responses from an estimated VAR-MIDAS methodology reveal that the EV of air carrier firms do not react uniformly to shocks to financial factors or economic and economic policy uncertainties.  One can only conclude that these firms have unique strategies to cope with financial, economic, and policy shocks.

The accumulated impulse responses indicate that most firms in the sample exhibit negative reactions to recessionary risks, as well as to domestic and global economic policy uncertainties.

Turning to the accumulated impulse responses of the sample carriers’ EV to financial shocks, most of the firms under study experience a positive accumulated response to positive shocks to their current ratio. The accumulated impulse response of the EV of all firms in the sample is negative in reaction to a positive one standard deviation shock to the debt to asset ratio, indicating that high air carrier firms' debt levels are not favored by the market. The accumulated response of the EV of many of the carriers in the sample show a positive reaction to their operating income after depreciation. The remainder of financial shocks do not have a clear accumulated impact on the EV of the firms under study.

In summary, the empirical findings in the present study confirm theoretical predictions in the literature. The existing literature on uncertainty shows that firms design optimal ways of coping with uncertainties. However, given that air carrier firms have disparate financial and market share positions, it is expected that their responses to the economic, policy, and financial uncertainties will not be uniform.

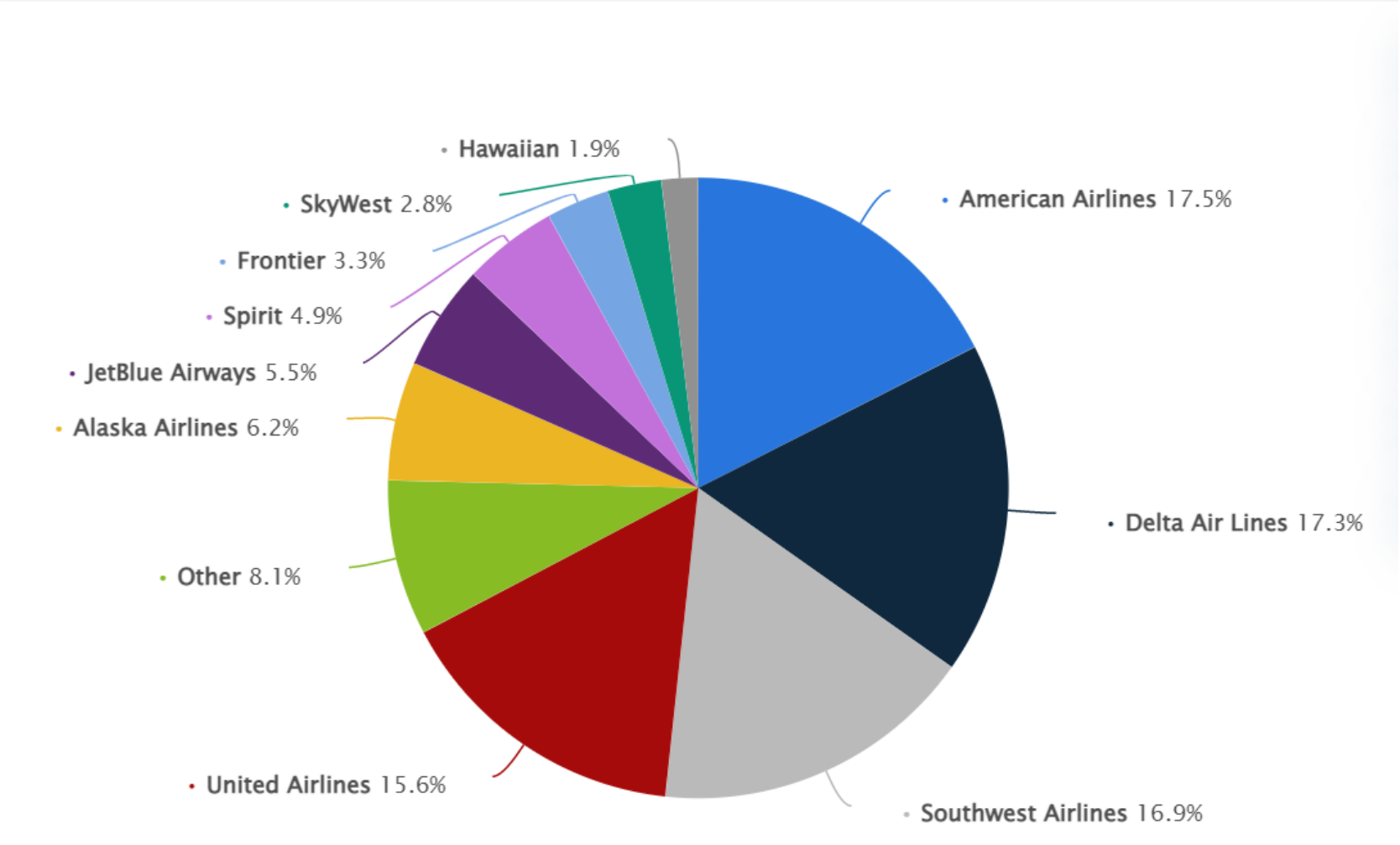

Figure 1: Domestic market share of leading U.S. airlines from February 2022 to January 2023
Source: Statista.com
https://www.statista.com/statistics/275948/market-capitalization-of-selected-airlines/

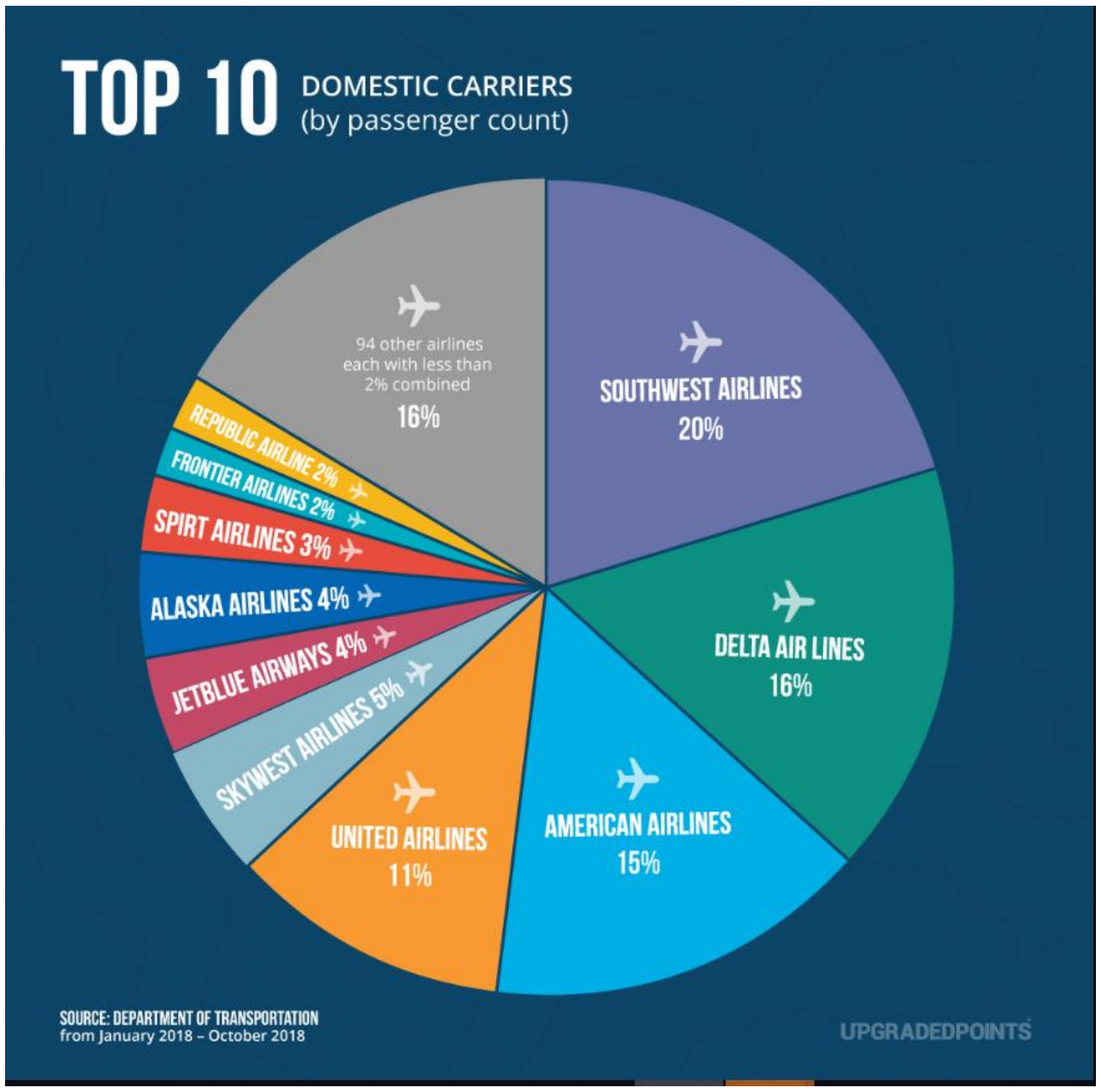

Figure 2: Domestic market share of leading U.S. airlines by passenger count in 2018.
Source: Department of transportation.

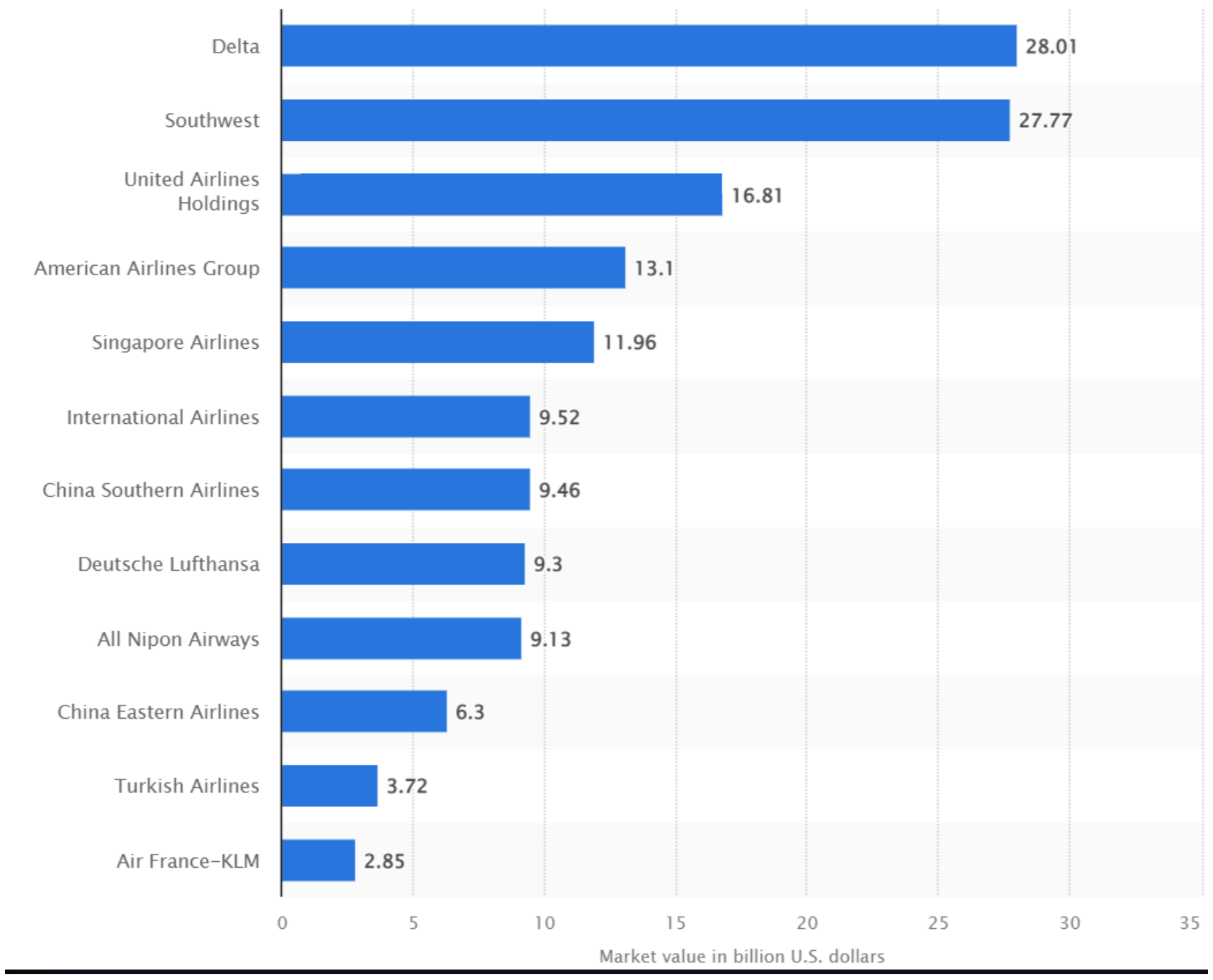

Figure 3: Market Value of the leading world airlines in 2022.
Source: Statista.com https://www.statista.com/statistics/275948/market-capitalization-of-selected-airlines/

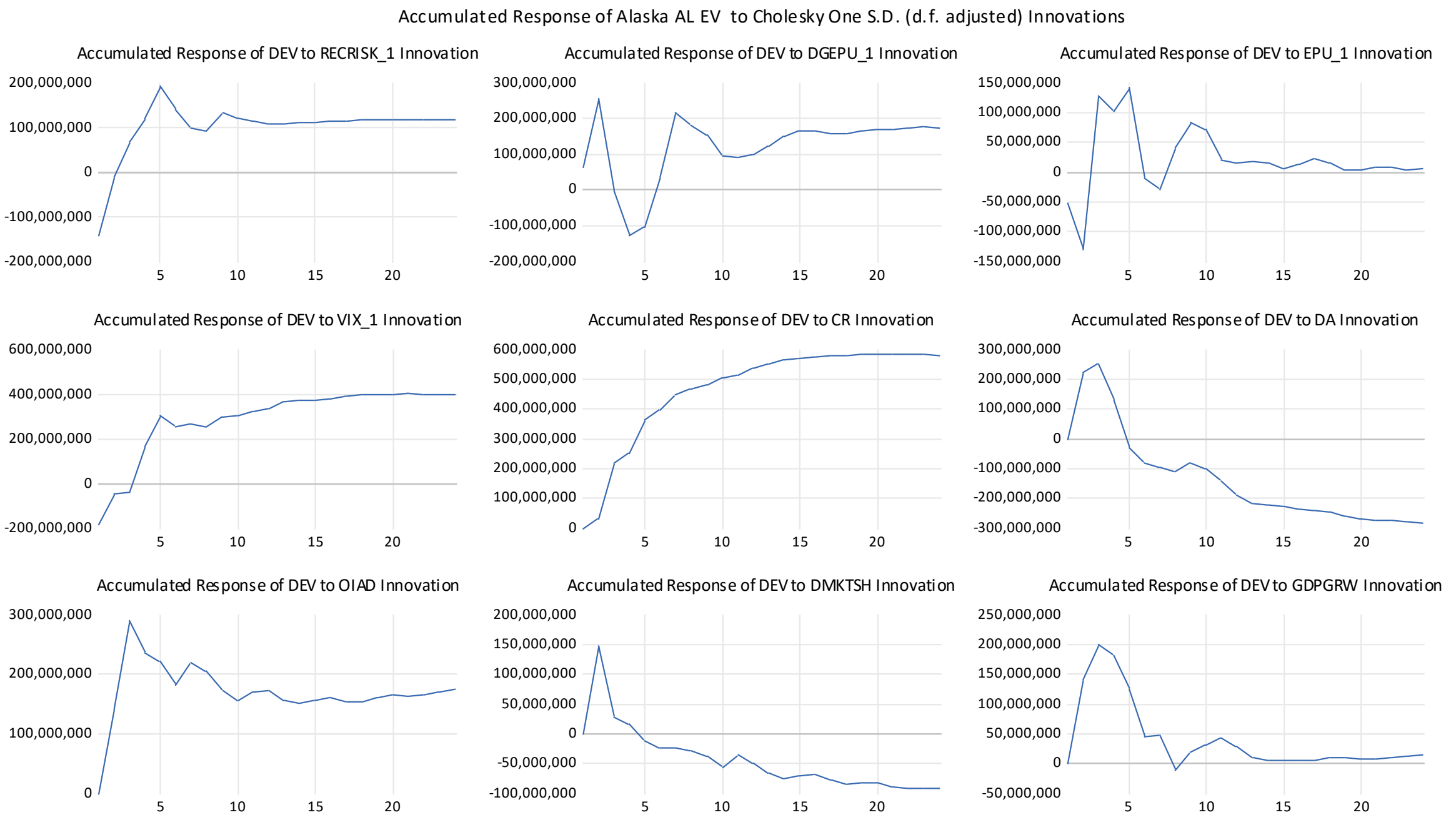

Figure 4: Accumulated impulse responses of the enterprise value of Alaska Airlines to a one standard deviation positive shock to uncertainty and financial indicators.

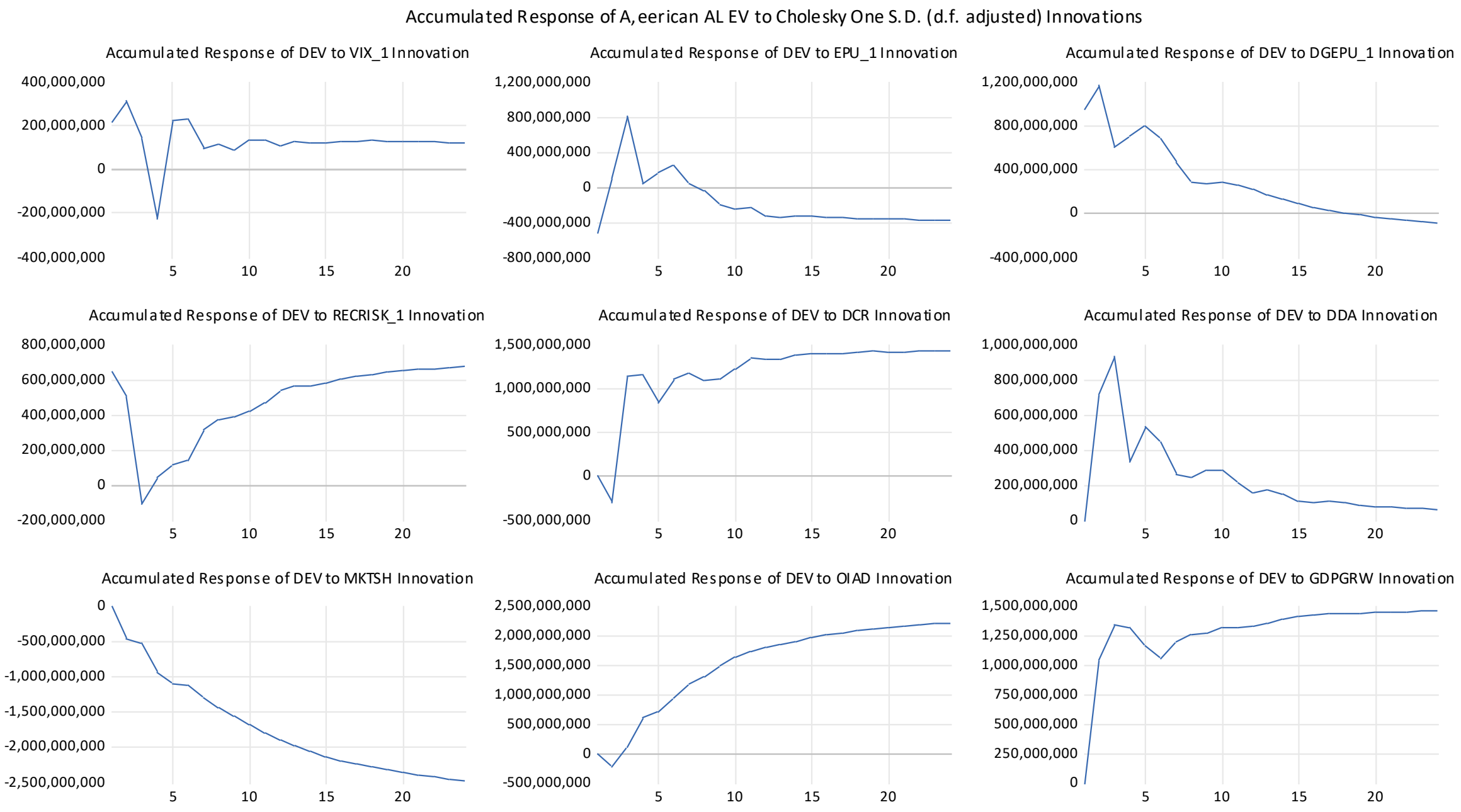

Figure 5: Accumulated impulse responses of the enterprise value of American Airlines to a one standard deviation positive shock to uncertainty and financial indicators.

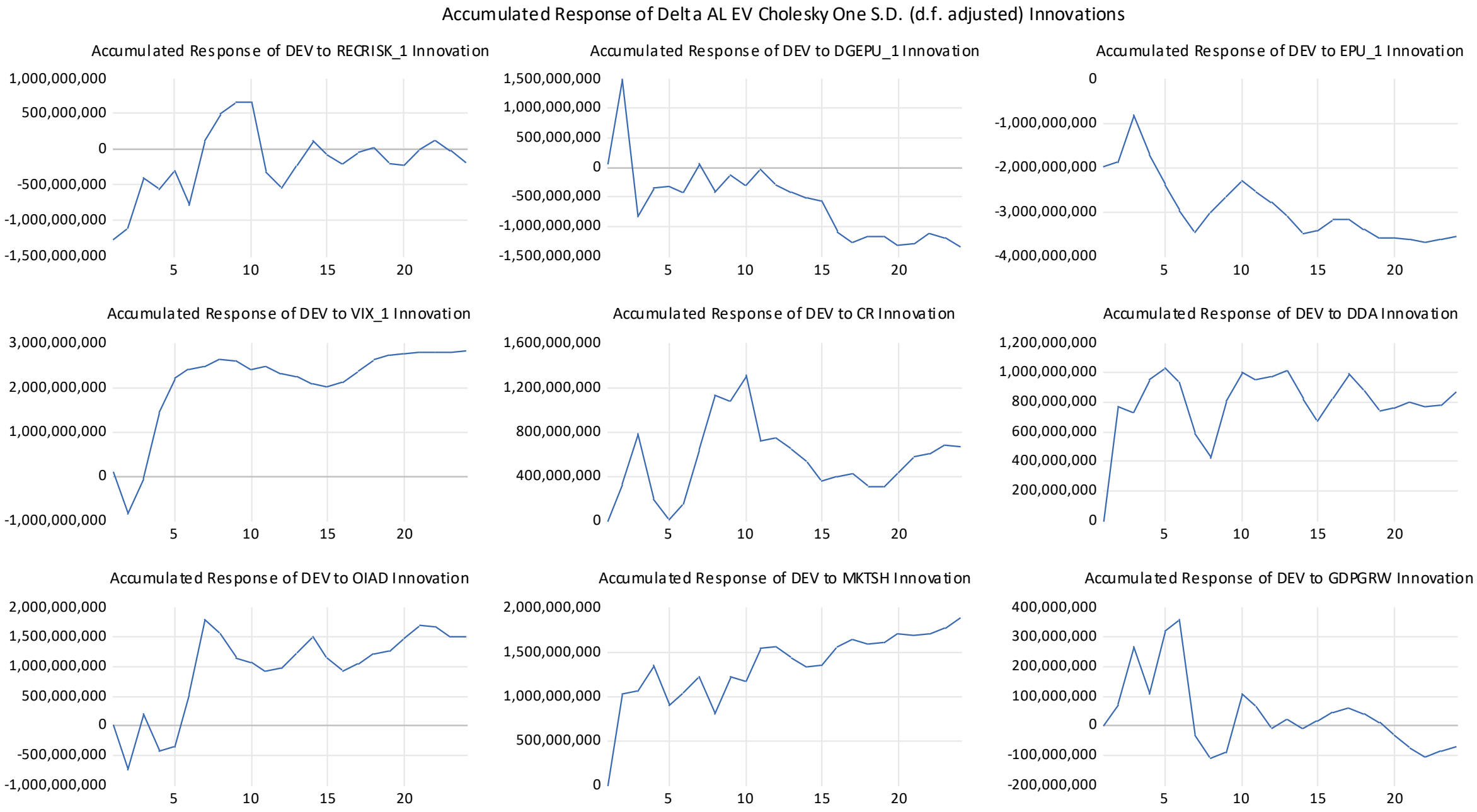

Figure 6: Accumulated impulse responses of the enterprise value of Delta Airlines to a one standard deviation positive shock to uncertainty and financial indicators.

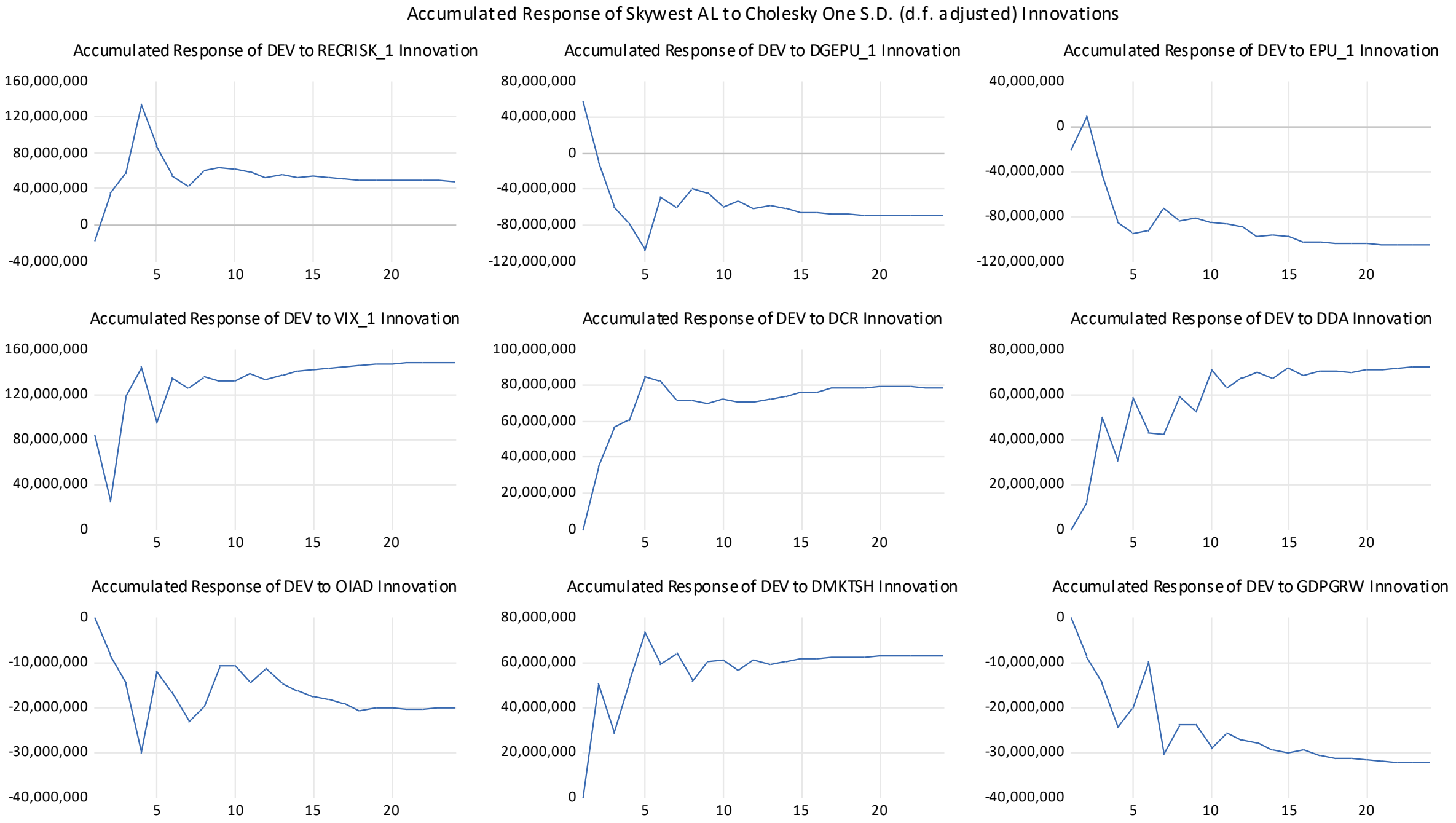

Figure 7: Accumulated impulse responses of the enterprise value of Skywest airlines to a one standard deviation positive shock to uncertainty and financial indicators.

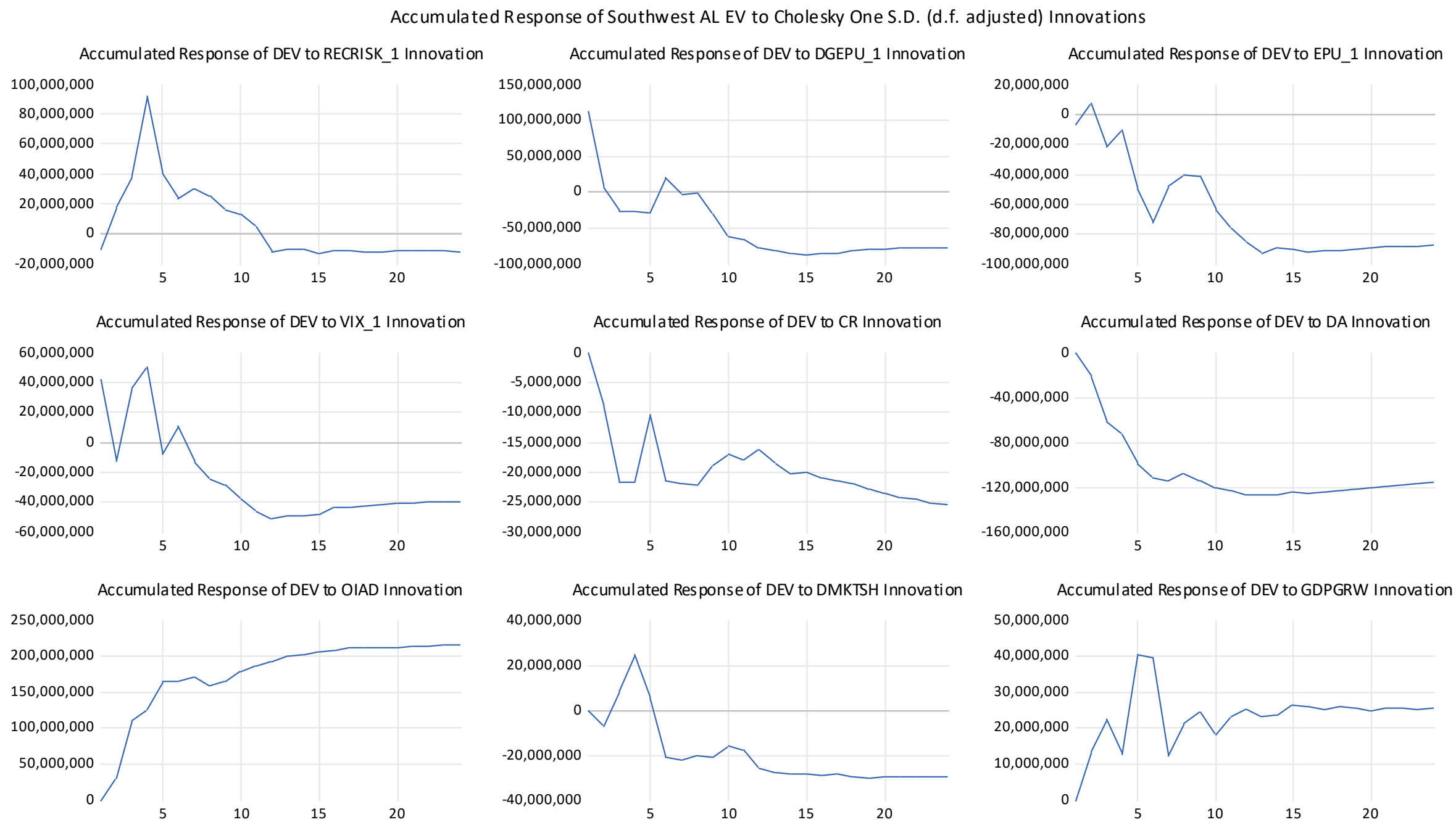

Figure 8: Accumulated impulse responses of the enterprise value of Southwest Airlines to a one standard deviation positive shock to uncertainty and financial indicators.

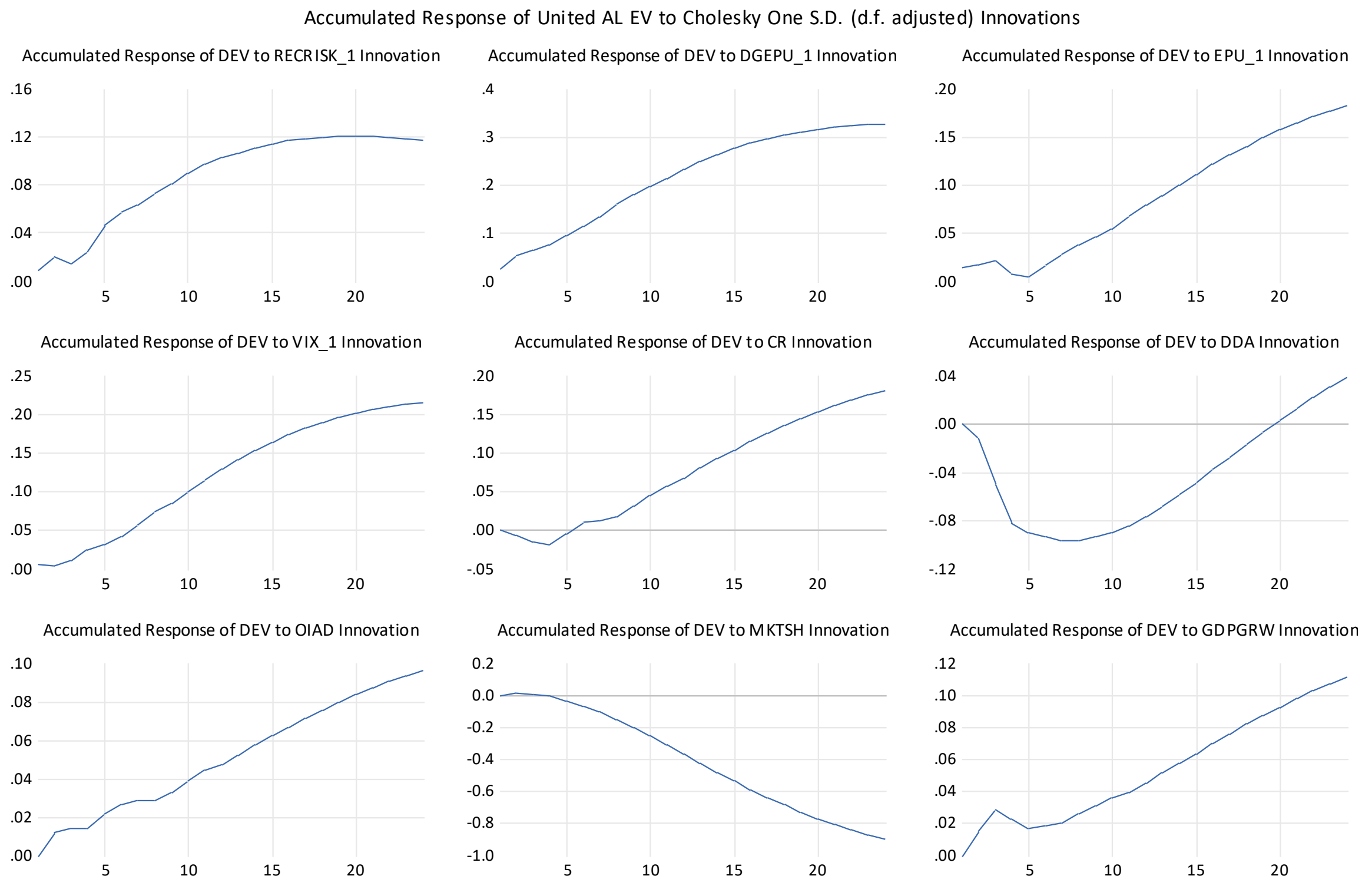

Figure 9: Accumulated impulse responses of the enterprise value of United Airlines to a one standard deviation positive shock to uncertainty and financial indicators.

Table 2: Mean of variables and Stationarity test results

Panel A: Mean

| AL | CR | DA | EV | MKtSH | OIAD |
|---|---|---|---|---|---|
| Alaska | 1.00 | 0.30 | 2.99*10^9 | 0.011 | 54.77 |
| American | 0.80 | 0.34 | 1.8*10^10 | 0.056 | 133.83 |
| Delta | 0.69 | 0.11 | 3.36*10^10 | 0.087 | 270.48 |
| United | 0.97 | 0.42 | 6.97*10^10 | 0.028 | 131.87 |
| SkyWest | 2.36 | 0.34 | 1.7*10^9 | 0.003 | 31.24 |
| Southwest | 1.11 | 0.22 | 8.35*10^8 | 0.007 | 172.73 |

Panel B: Augmented Dickey Fuller and Phillips-Perron Stationarity test Results

| | | | | | |
|---|---|---|---|---|---|
| Alaska | | | | | |
| ADF | -4.408a | -2.958b | -2.062 | -2.094 | -3.328b |
| PP | -4.301a | -2.820c | 1.421 | -4.631a | -5.053a |
| American | | | | | |
| ADF | -1.861 | 0.7766 | -1.329 | -3.066b | -3.344b |
| PP | -1.861 | -1.450 | -1.299 | -3.066b | -4.081a |
| Delta | | | | | |
| ADF | -3.620a | -2.179 | -1.971 | -5.463a | -3.313b |
| PP | -3.182a | -2.385 | -1.733 | -8.357a | -4.931a |
| United | | | | | |
| ADF | -3.644a | -2.677 | -2.197 | -3.358a | -4.030a |
| PP | -4.083a | -2.576 | -2.079 | -3.743a | -4.572a |
| SkyWest | | | | | |
| ADF | -2.545c | -1.939 | -0.969 | -2.471 | -4.095 |
| PP | -2.531 | -1.786 | -0.919 | -2.419 | -4.057 |
| Southwest | | | | | |
| ADF | -4.144a | -3.729a | -1.895 | -0.557 | -4.594a |
| PP | -4.100a | -4.155a | -1.750 | -1.178 | -4.673a |
| Uncertainty Index | EPU | GEPU | Recession Risk | VIX | |
| ADF | -8.447a | -1.756 | -1.895 | -6.243a | |
| PP | -8.603a | -2.864a | -1.750 | -5.238a | |

Notes: Enterprise values of United AL, American AL, and Delta AL are in billions of dollars. Those of the remaining airlines are in millions of dollars. If a variable for any airlines shows unit root, that variable enters VARs in first difference.

Table 3: Direction of the Accumulated Impulse Response of the Enterprise Value to Shocks to Financial Variables

| AL | CR | DA | MKTSH | OIAD |
|---|---|---|---|---|
| Alaska | + | - | Mixed | + |
| American | + | - | - | + |
| Delta | Mixed | Mixed | + | + |
| SkyWest | + | + | + | Mixed |
| Southwest | - | - | Mixed | + |
| United | + | - | - | + |

Table 4: Direction of the Accumulated Impulse Response of the Enterprise Value to Shocks to Uncertainty Indices

| AL | VIX | Recession Risk | GEPU | EPU |
|---|---|---|---|---|
| Alaska | + | None | + | Mixed |
| American | Mixed | - | - | - |
| Delta | + | - or Mixed | - | - |
| SkyWest | + | None | - | - |
| Southwest | Mixed | - | - | - |
| United | + | + | Mixed | + |